\documentclass[conference]{IEEEtran}
\IEEEoverridecommandlockouts
\usepackage{textcomp}
\usepackage{xcolor}
\usepackage{amsmath,amssymb,amsfonts}
\usepackage{graphicx}
\usepackage[hidelinks]{hyperref}
\usepackage{algpseudocode}
\usepackage{algorithm}
\usepackage{multirow}
\usepackage{booktabs} 
\usepackage{tabularx, booktabs}
\usepackage{threeparttable}
\usepackage{subfigure}
\usepackage{pifont}
\usepackage[numbers,sort&compress]{natbib}
\usepackage{pifont} 
\newcommand{\cmark}{\ding{51}}
\newcommand{\xmark}{\ding{55}}
\usepackage{bbding}

\def\BibTeX{{\rm B\kern-.05em{\sc i\kern-.025em b}\kern-.08em
    T\kern-.1667em\lower.7ex\hbox{E}\kern-.125emX}}

\begin{document}
\title{MEEG and AT-DGNN: Improving EEG Emotion Recognition with Music Introducing and Graph-based Learning}

\author{
    \IEEEauthorblockN{1\textsuperscript{st} Minghao Xiao}
    \IEEEauthorblockA{\textit{Shenzhen Research Institute} \\
    \textit{of Shandong University} \\
    Shenzhen, China \\
    xmh001011@gmail.com}
    \and
    \IEEEauthorblockN{1\textsuperscript{st} Zhengxi Zhu}
    \IEEEauthorblockA{\textit{Shenzhen Research Institute} \\
    \textit{of Shandong University} \\
    Shenzhen, China \\
    202200800179@mail.sdu.edu.cn}
    \and
    \IEEEauthorblockN{3\textsuperscript{rd} Kang Xie*}
    \IEEEauthorblockA{\textit{Key Lab of Information} \\
    \textit{Network Security} \\
    \textit{Ministry of Public Security} \\
    Shanghai, China \\
    xiekang@stars.org.cn}
    \and
    \IEEEauthorblockN{4\textsuperscript{th} Bin Jiang}
    \IEEEauthorblockA{\textit{Shenzhen Research Institute} \\
    \textit{of Shandong University} \\
    Shenzhen, China \\
    jiangbin@sdu.edu.cn}
}

\maketitle

\begin{abstract}

We present the MEEG dataset, a multi-modal collection of music-induced electroencephalogram (EEG) recordings designed to capture emotional responses to various musical stimuli across different valence and arousal levels. This public dataset facilitates an in-depth examination of brainwave patterns within musical contexts, providing a robust foundation for studying brain network topology during emotional processing. Leveraging the MEEG dataset, we introduce the Attention-based Temporal Learner with Dynamic Graph Neural Network (AT-DGNN), a novel framework for EEG-based emotion recognition. This model combines an attention mechanism with a dynamic graph neural network (DGNN) to capture intricate EEG dynamics. The AT-DGNN achieves state-of-the-art (SOTA) performance with an accuracy of 83.74\% in arousal recognition and 86.01\% in valence recognition, outperforming existing SOTA methods. This study advances graph-based learning methodology in brain-computer interfaces (BCI), significantly improving the accuracy of EEG-based emotion recognition. The MEEG dataset and source code are publicly available at \textit{https://github.com/xmh1011/AT-DGNN}.

\end{abstract}


\IEEEpeerreviewmaketitle

\section{Introduction}

Emotion plays a key role in interpreting human responses, and Russell’s Valence-Arousal (VA) model offers a dimensional framework for quantifying emotional states within a continuous space \cite{russell1980circumplex}. Meanwhile, BCI technology, which decodes EEG signals reflecting cortical voltage fluctuations, translates thoughts into commands for external devices, significantly enhancing human-computer interaction \cite{zhao2021emotion}. With their high temporal resolution, objectivity, affordability, and quick acquisition, EEG signals are crucial for emotion recognition and neuroscience research, particularly in integrating emotional models with advanced decoding methods \cite{foong2019assessment}.

In 2011, Koelstra et al. introduced the DEAP dataset, consisting of physiological signals like EEG elicited by emotional responses to musical videos \cite{koelstra2011deap}, followed by Zheng et al.'s SEED dataset in 2015, which provided well-annotated EEG signals from cinematic stimuli across multiple subjects \cite{zheng2015investigating}. These datasets have greatly advanced computational models for emotion analysis. 

Recent advances in deep learning have significantly impacted EEG signal processing. In 2018, Lawhern et al. proposed EEGNet, a compact convolutional neural network (CNN) tailored for EEG analysis \cite{lawhern2018eegnet}. Subsequent developments by Schirrmeister et al. further refined EEG decoding capabilities using deep CNN architectures, introducing sophisticated visualization tools \cite{schirrmeister2017deep}. The adoption of temporal convolutional networks (TCNs) has introduced robust alternatives for real-time BCI applications. Notably, in 2020, Ingolfsson et al. developed EEG-TCNet, which demonstrated superior accuracy in motor imagery tasks, suggesting that TCNs could outperform traditional methods in specific BCI contexts \cite{ingolfsson2020eeg}. This assertion was corroborated by Musallam et al., who demonstrated the versatility and efficiency of TCNs in complex BCI tasks, particularly through the integration of TCNs in motor imagery classification \cite{musallam2021electroencephalography}.

All aforementioned studies  treated EEG signals as two-dimensional time-series data, with channels representing EEG electrodes positioned according to the 10-20 system to capture neural activity across brain regions \cite{can2023approaches, zhao2021emotion}. Recently, EEG data have been increasingly modeled as graphs, representing the spatial arrangement of electrodes, where features from each electrode add a third dimension to the model \cite{grana2023review, robinson2019eeg}.

In 2019, Song et al. introduced a novel DGCNN for multichannel EEG emotion recognition, utilizing an adjacency matrix to dynamically model EEG channel relationships and enhance feature discrimination, demonstrating superiority over existing approaches  \cite{song2018eeg}. Subsequently, advancements in GNNs have been increasingly applied to emotion recognition. Bao et al. integrated a multi-layer GNN with a style-reconfigurable CNN \cite{bao2022linking}, while Asadzadeh et al. further improved the DGCNN by incorporating Bayesian signal recovery techniques, both achieving enhanced performance \cite{asadzadeh2022accurate}. 

Building on prior advancements, greater progress was made in 2023 when Ding et al. introduced the Local-Global-Graph Network (LGGNet) for processing EEG data in an image-like format, achieving improved classification performance \cite{ding2023lggnet}. Alternatively, LGGNet's single-layer graph neural network faces challenges in capturing the dynamic features of EEG signals. Additionally, widely used EEG datasets such as DEAP and SEED primarily involve visual stimuli, which limits exploration of the relationship between music, emotions, and EEG signals. Conversely, EEG datasets and research using musical stimuli are relatively scarce. Moreover, existing SOTA methods achieve lower accuracy on the DEAP dataset, thus limiting comprehensive performance evaluation of models.

To address these limitations in emotion recognition research, the MEEG (Music EEG) dataset is developed accordingly. This multi-modal dataset, similar to DEAP, utilizes music-induced emotional states to improve the accuracy of emotion recognition. By integrating a sliding window technique, an attention mechanism, and a multi-layer dynamic graph neural network into the LGGNet architecture, significant improvements in classification accuracy are achieved. 

The contributions of our study are threefold and can be summarized as follows:

\begin{enumerate}
    \item The MEEG dataset, a multi-modal EEG emotion dataset in the DEAP format, is enhanced with diverse music to induce emotional states effectively. It outperforms the DEAP dataset in emotion induction, improving model accuracy (ACC) and F1 scores.
    \item The AT-DGNN framework explores connections within and between brain functional areas. By integrating a sliding window technique, an attention mechanism, and a stacked DGNN, this novel architecture for emotion recognition enhances the learning of dynamic features in EEG recordings.
    \item Performance of the AT-DGNN is compared with other CNN, TCN, and GNN-based SOTA methods on MEEG dataset. Ablation experiments provide insights into the AT-DGNN framework.
\end{enumerate}

\section{Related Work}

\begin{figure*}[htbp]
    \centering
    \includegraphics[width=\linewidth]{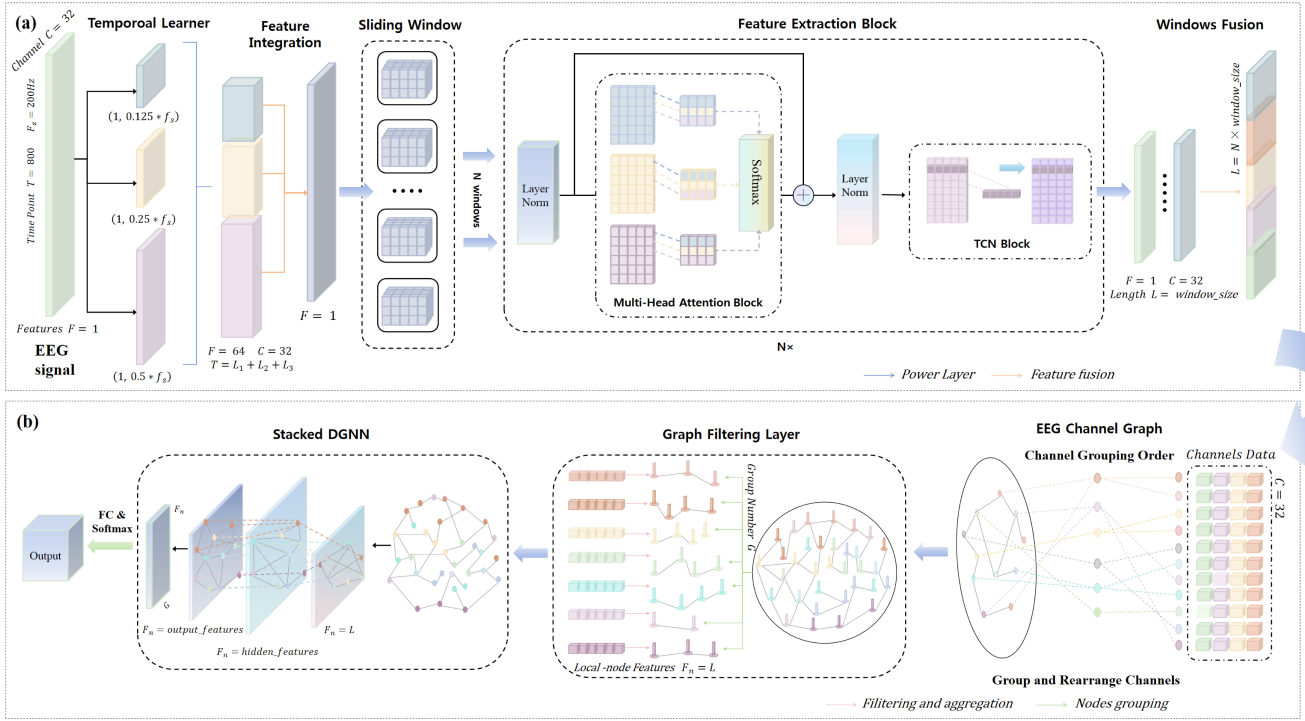}
    \makeatletter
    \def\@IEEEfigurecaptionsepspace{\vskip\abovecaptionskip}
    \def\@IEEEfigurecaptionspace{\vskip\belowcaptionskip}
    \def\@IEEEeqnarrayfigurecaptionspace{\vskip 3pt}
    \long\def\@makecaption#1#2{\@IEEEfigurecaptionsepspace\footnotesize\hbox to \hsize{\hfil\parbox[t]{\hsize}{#1: #2}\hfil}\@IEEEfigurecaptionspace}
    \makeatother
    \caption{Structure of AT-DGNN. The AT-DGNN model comprises two core modules: a feature extraction module (a) and a dynamic graph neural network learning module (b). The feature extraction module consists of a temporal learner, a multi-head attention mechanism, and a temporal convolution module. These components effectively leverage local features of EEG signals through a sliding window technique, thereby enhancing the model's capacity to dynamically extract complex temporal patterns in EEG signals. In the graph-based learning module, the model initially employs local filtering layers to segment and filter features from specific brain regions. Subsequently, the architecture employs three layers of stacked dynamic graph convolutions to capture complex interactions among different brain regions. This structure enhances the AT-DGNN's capacity for integrating temporal features effectively.}
    \label{fig:1}
\end{figure*}

\subsection{Multi-head Attention Mechanism}

EEG data is inherently complex, with significant temporal dependencies and discrete features. Multi-head attention (MHA) helps capture diverse temporal patterns and dependencies, enabling the identification of key features. In this module, the attention block consists of a multi-head attention layer with several self-attention heads. As shown by Zhang et al. \cite{ZHANG2023119107}, multi-head self-attention effectively integrates multi-source information, improving the model’s robustness in handling complex temporal features in EEG data, leading to enhanced classification performance.

Each self-attention head comprises three fundamental components: queries \( Q \), keys \( K \), and values \( V \). These elements facilitate the computation of attention scores that influence the weighting of the values. The process begins with normalizing the input \( X_w \) via a layer normalization (\(\textit{LayerNorm}\)):

\begin{equation}
\begin{cases}
q_{ht} = W^Q \textit{LayerNorm}(X_{w,t}) \\
k_{ht} = W^K \textit{LayerNorm}(X_{w,t}) \\
v_{ht} = W^V \textit{LayerNorm}(X_{w,t})
\label{eq:1}
\end{cases}
\end{equation}

Here, \( q_{ht} \), \( k_{ht} \), and \( v_{ht} \) denote the queries, keys, and values at time \( t \) for head \( h \), derived from the normalized input. The matrices \( W^Q \), \( W^K \), and \( W^V \) belong to \(\mathbb{R}^{d \times d_H}\), where \( d_H \) represents the dimension of each attention head. This normalization facilitates the efficient computation of attention scores.

The attention context vector \( c_{ht} \) for each head is computed as a weighted sum of the values, with weights \( \alpha_{htt'} \) determined by the scaled dot-product attention mechanism. The alignment scores \( e_{htt'} \) are calculated as:

\begin{equation}
e_{htt'} = \frac{(q_{ht})^T k_{ht'}}{\sqrt{d_H}}
\label{eq:2}
\end{equation}

Here, \( t' \) denotes the different time steps considered. The softmax function converts these scores into a probability distribution, reflecting the importance of each value vector:
\begin{equation}
c_{ht} = \sum_{t'} \alpha_{htt'} v_{ht'}
\label{eq:3}
\end{equation}
where \( \alpha_{htt'} \) represents the normalized alignment scores, indicating the importance of each value vector \( v_{ht'} \) in forming the context vector \( c_{ht} \) for time step \( t \). This context vector \( c_{ht} \) captures the most relevant information from the values, highlighting key features at each time step.

\subsection{Graph Neural Networks}

GNNs represent a significant advancement in the domain of neural networks, designed explicitly to process graph-structured data. Unlike CNNs, GNNs excel in capturing the intricate relationships and dependencies among nodes within a graph through processes of aggregation and propagation of information across local neighbors \cite{scarselli2008graph}. A typical graph is denoted as \( G = (V, E) \), where \( V \) symbolizes the set of nodes and \( E \) the set of edges. Each node \( v_i \in V \) and edge \( e_{ij} = (v_i, v_j) \in E \) can be respectively associated with a node and an edge in the graph. The adjacency matrix \( A \) is configured as an \( n \times n \) matrix where \( A_{ij} = 1 \) if \( e_{ij} \in E \) and \( A_{ij} = 0 \) otherwise. Node attributes are represented by \( X \), where \( X \in \mathbb{R}^{n \times d} \), with each \( x_i \in \mathbb{R}^d \) denoting the feature vector of node \( v_i \).

DGNNs extend the capabilities of GNNs to address dynamic or time-evolving graph-structured data. In DGNNs, a graph at any given time \( t \) is represented as \( G_t = (V_t, E_t) \), with its corresponding adjacency matrix \( A_t \). This matrix changes dynamically as nodes and edges are added or removed over time. Node features at time \( t \) are likewise dynamic, represented as \( X_t \) with each row \( x_{t,i} \in \mathbb{R}^d \) embodying the evolving feature set of node \( v_i \).

The computational heart of DGNNs lies in the dynamic update rules, where node representations \( h_{t,i} \) are recurrently updated based on the temporal graph structure. A prevalent approach involves the temporal graph attention mechanism, where the new node states are computed as:

\begin{equation}
\textit{h}_{t,i}^{(k+1)} = \sigma\left(\sum_{j \in \mathcal{N}(v_i)} \alpha_{t,ij}^{(k)} W^{(k)} \textit{h}_{t,j}^{(k)} + \textit{b}^{(k)}\right)
\label{eq:4}
\end{equation}

Here, \( \mathcal{N}(v_i) \) represents the neighbors of node \( v_i \), \( \alpha_{t,ij}^{(k)} \) are the attention coefficients indicating the significance of the features of neighbor \( j \) to node \( i \), and \( W^{(k)} \) and \( b^{(k)} \) are trainable parameters of the \( k \)-th layer, with \( \sigma (\cdot)\) being a non-linear activation function.

\section{Methodology}

In this section, we introduce the detailed architecture of AT-DGNN from the three aspects of data preprocessing, feature extraction and graph-based learning in turn, and the overall structure of AT-DGNN is shown in Fig.~\ref{fig:1}.

\subsection{EEG Data Preprocessing}

The initial sampling rate of the MEEG dataset is 1000 Hz. To align with standard datasets typically sampled at 128 Hz or 200 Hz, and to mitigate issues such as high artifact noise and overlapping interference signals, the data are downsampled to 200 Hz. This preprocessing phase involves the use of a band-pass filter with a range of 1 to 50 Hz to enhance the signal quality by minimizing interference. Subsequently, the EEG signals are segmented into five distinct frequency bands for feature extraction: Delta (1-4 Hz), Theta (4-8 Hz), Alpha (8-14 Hz), Beta (14-31 Hz), and Gamma (31-50 Hz).

\subsection{Feature Extraction}

\subsubsection{Temporal Learner}

A temporal learner layer employing multiscale 1D temporal kernels ($T$ kernels) is used to directly extract dynamic temporal representations from EEG data \( X_{i} \in \mathbb{R}^{E \times T} \), where \( E \) denotes the number of EEG electrodes, and \( T \) is the sample length. These kernels obviate the need for manually extracted features. The \(i\)-th kernel's length (\(S_{T}^{i}\)), dictated by the sampling frequency ($f_{s}$) of EEG data and scaling coefficients ($\alpha^i$), can be represented by \cite{ding2020tsception}:

\begin{equation}
S_{T}^{i} = (1, \alpha^{i} \cdot f_{s}), \quad i \in [1, 2, 3]
\label{eq:5}
\end{equation}

EEG data processed through these layers yield dynamic time-frequency representations. An average pooling (\textit{AvgPool}) layer, acting as a window function, calculates the averaged power across shorter segments. The logarithmic activation as described by \cite{schirrmeister2017deep} is applied to improve the performance. The output from each layer \(i\), denoted as \( Z^{i}_{\textit{temp}} \), is formulated as:

\begin{equation}
Z^{i}_{\textit{temp}} = \Phi_{\textit{log}}(\textit{AvgPool}(\Phi_{\textit{sqr}}(\mathcal{F}_{\textit{Conv1-D}}(X_{i}, S^{i}_{T}))))
\label{eq:6}
\end{equation}

Here, \( \Phi_{\textit{log}}(\cdot) \) is the logarithmic activation function, \( \Phi_{\textit{sqr}}(\cdot) \) represents the square function, and \( \mathcal{F}_{\textit{Conv1-D}}(\cdot) \) signifies the 1D convolution operation.

The outputs across all kernels are concatenated along the feature dimension to yield the final output of the temporal learner layer \( Z_{T} \):
\begin{equation}
Z_{T} = f_{\textit{bn}}(Z^{1}_{\textit{temp}}, Z^{2}_{\textit{temp}}, Z^{3}_{\textit{temp}})
\label{eq:7}
\end{equation}
where \( f_{\textit{bn}} (\cdot)\) denotes the batch normalization operation.

\subsubsection{Sliding Window Segmentation}

Following the temporal learner, a convolution-based sliding window technique as described in \cite{altaheri2022physics} is applied to segment the EEG time series $X$ into multiple windows. This method integrates sliding window segmentation with convolutional operations, significantly reducing computational overhead by allowing convolutions to be executed once across all windows. This approach enhances data augmentation while accelerating processing through parallel computation. The time series $X$ is segmented into windows $X_w \in \mathbb{R}^{B \times C \times W}$ using a window of length $W$ and stride $S$, where $w = 1, \ldots, n$ represents the window index and $n$ is the total number of windows. Each window $X_w$ is then processed by subsequent attention and temporal convolution blocks. The number of windows $n$ is calculated as:

\begin{equation}
n = \left\lfloor \frac{\textit{length} - \textit{window\_size}}{\textit{stride}} \right\rfloor + 1
\label{eq:8}
\end{equation}

Here, $\textit{length}$, $\textit{window\_size}$, and $\textit{stride}$ denote the total length of the time series, the length of each window, and the stride between windows, respectively.

\subsubsection{Multi-head Attention Module}

Each subsequence \(X_w\) derived from sliding window segmentation is processed as described in \eqref{eq:1}. The attention context vector \(c_{ht}\) for each head is computed as a weighted sum of the values, with weights \(\alpha_{htt'}\) determined by the scaled dot-product attention mechanism as \eqref{eq:2}, and then used in \eqref{eq:3}.

The context vectors \(c_{1t}, \ldots, c_{Ht}\) from all heads are concatenated and linearly transformed to yield the final output of the MHA layer:
\begin{equation}
\textit{X}_w' = W^O [\textit{c}_{1t}, \ldots, \textit{c}_{Ht}] + \textit{X}_w,
\label{eq:9}
\end{equation}
where \(W^O \in \mathbb{R}^{d_H \times d}\) is the output projection matrix. This procedure projects the combined outputs back to the original input dimension. Subsequently, \(X_w'\) is integrated with the original subsequence \(X_w\) through a residual connection and normalized once more. This advanced MHA framework significantly enhances the model's capacity to discern intricate temporal patterns and dependencies within EEG signals, thereby augmenting the precision of decoding activities.

\subsubsection{Temporal Convolution}

Following the MHA mechanism, the output \( \textit{X}_w' \) is normalized using \(\textit{LayerNorm}\). The normalized EEG data are denoted as \(X_i \in \mathbb{R}^{c \times l}\), where \(c\) represents the number of EEG channels and \(l\) represents the temporal dimension's sample length.

To extract dynamic temporal features, a temporal convolutional block is applied to each window of the EEG data. The normalized EEG data \(X_i\) serves as the input to this block. The temporal convolutional block processes \(X_i\) to produce an output tensor \(Z_{\textit{tcn}}^w \in \mathbb{R}^{B \times C \times T_w}\), where \(T_w\) is the temporal length of the window and \(w\) is the window index. Batch normalization is applied to ensure training stability, followed by a \(\textit{ReLU}(\cdot)\) activation function to introduce non-linearity. The output \(Z_{\textit{tcn}}^w\) is computed as follows:

\begin{equation}
Z_{\textit{tcn}}^w = \textit{ReLU}(\textit{BatchNorm}(\textit{Conv1d}(X_i)))
\label{eq:10}
\end{equation}

\subsubsection{Feature Fusion}

The outputs of the temporal learner for each window, denoted as \(Z^w_{\textit{tcn}} \in \mathbb{R}^{B \times C \times T_w}\), are assembled into a four-dimensional tensor \(Z_{\textit{stacked}}\):

\begin{equation}
Z_{\textit{stacked}} = [Z^1_{\textit{tcn}}, Z^2_{\textit{tcn}}, \ldots, Z^W_{\textit{tcn}}] \in \mathbb{R}^{B \times C \times W \times T_w}
\label{eq:11}
\end{equation}

This tensor is rearranged and flattened to meet the convolution layer's input requirements, resulting in dimensions \((B, 32, -1)\), where \(32\) is the fixed channel count, and \(-1\) represents the flattened dimensions. A fusion convolution layer with a kernel size of 3 and a stride of 1 integrates the outputs across all windows. The final output \(Z_{\text{fused}} \in \mathbb{R}^{B \times 32 \times L}\), where \(L\) is the combined length, captures global contextual information, allowing for a more discriminative representation of features in EEG signals, which is crucial for constructing nodes in GNNs.

\subsection{Graph-based Learning} 

\subsubsection{Graph Filtering Layer}

\begin{figure}[ht]
    \centering
    \includegraphics[width=\linewidth]{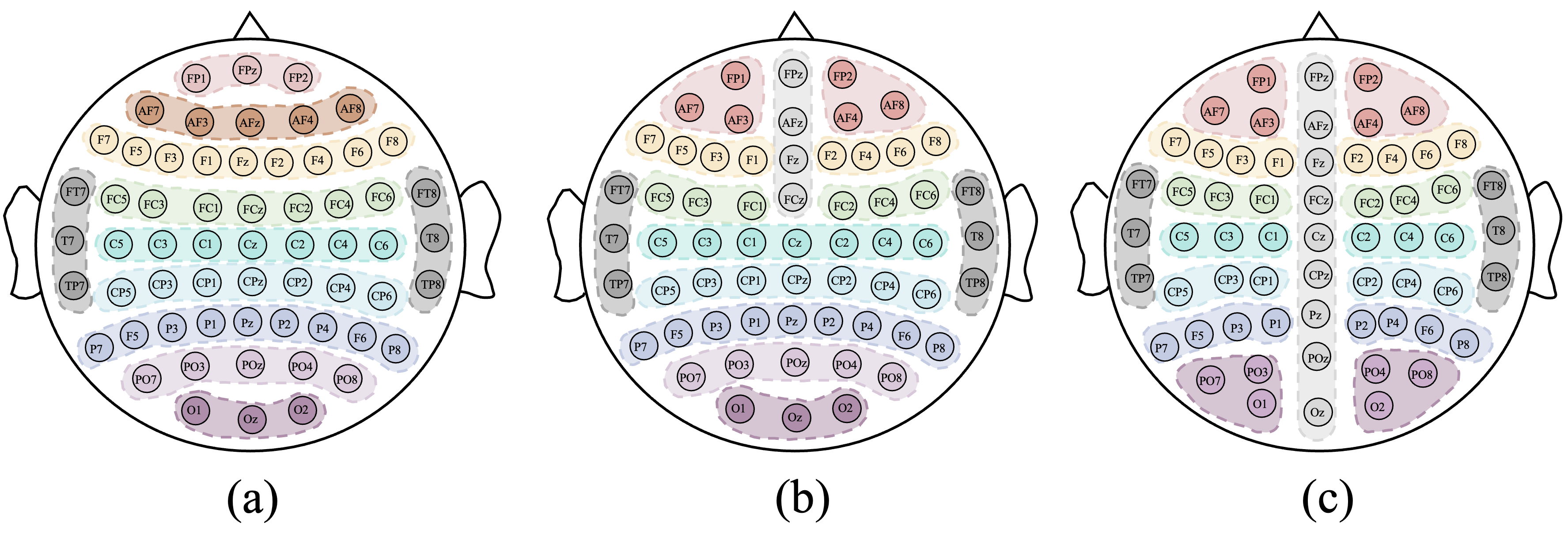}
    \makeatletter
    \def\@IEEEfigurecaptionsepspace{\vskip\abovecaptionskip}
    \def\@IEEEfigurecaptionspace{\vskip\belowcaptionskip}
    \def\@IEEEeqnarrayfigurecaptionspace{\vskip 3pt}
    \long\def\@makecaption#1#2{\@IEEEfigurecaptionsepspace\footnotesize\hbox to \hsize{\hfil\parbox[t]{\hsize}{#1: #2}\hfil}\@IEEEfigurecaptionspace}
    \makeatother
    \caption{Local-global graph definitions \cite{ding2023lggnet}. (a) General definition $G_g$. (b) Frontal definition $G_f$. (c) Hemispheric definition $G_h$.}
    \label{fig:2}
\end{figure}

The method described by \cite{ding2023lggnet} is adopted for the extracted features, involving the definition of functional areas and local filtering. Electrodes are divided into three functional areas: the general region (\(G_g\)), the frontal region (\(G_f\)), and the hemispheric region (\(G_h\)), as shown in Fig.~\ref{fig:2}. Each channel is treated as a node, with the learned dynamic temporal representations considered as node attributes.

In the preprocessing step, EEG channels are systematically reordered within predefined groups to ensure adjacency of channels within each local graph. This reordering enhances the effectiveness of localized graph-based operations applied in subsequent stages. Mathematically, the reordering of channel representations, denoted by \(Z_{\textit{reorder}}^i\), is defined by the function \(F_{\textit{reorder}}(\cdot)\), which operates on the fused data from EEG channels at the \(i\)-th observation representations \(Z_{\textit{fuse}}^i\). This relationship is expressed as:
\begin{equation}
Z_{\textit{reorder}}^i = F_{\textit{reorder}}(Z_{\textit{fuse}}^i)
\label{eq:12}
\end{equation}

After reordering, a graph filtering layer is employed to aggregate the transformed representations of EEG channels using a fully connected local adjacency matrix \(A_{\textit{local}}\), where all elements are set to 1. The filtered output \(Z_{\textit{filtered}}^i\) is obtained by applying a trainable filter matrix \(W_{\textit{local}}\) and a bias \(b_{\textit{local}}\) to the reordered data, followed by the application of the \(\textit{ReLU}(\cdot)\) activation function:
\begin{equation}
Z_{\textit{filtered}}^i = \textit{ReLU}(W_{\textit{local}} \circ Z_{\textit{reorder}}^i + b_{\textit{local}})
\label{eq:13}
\end{equation}

The element-wise product symbol \(\circ\) captures localized interactions within the brain.

The output of the filtering process, \( Z_{\textit{local}}^i \), is aggregated to form localized feature representations using the aggregation function \( F_{\textit{aggregate}}(\cdot) \). The vector representation for each local graph is:
\begin{equation}
Z_{\textit{local}}^i = F_{\textit{aggregate}}(Z_{\textit{filtered}}^i) = 
\begin{bmatrix} 
h_{\textit{local}}^1 \\ 
\vdots \\ 
h_{\textit{local}}^R 
\end{bmatrix}
\label{eq:14}
\end{equation}
where \( h_{\textit{local}}^1, \dots, h_{\textit{local}}^R \) are elements of the localized feature representation vector. These features \( Z_{\textit{local}}^i \) are used to construct the overall feature matrix \( Z_{\textit{agg}} \):

\begin{equation}
Z_{\textit{agg}} = 
\begin{bmatrix} 
Z_{\textit{local}}^1 \\ 
Z_{\textit{local}}^2 \\ 
\vdots \\ 
Z_{\textit{local}}^{11} 
\end{bmatrix}
\label{eq:15}
\end{equation}

\subsubsection{Stacked DGNN}

A stacked Dynamic Graph Neural Network (DGNN) captures complex graph relationships by recalculating the adjacency matrix at each layer based on input features, effectively capturing hierarchical feature variations.

The architecture incorporates multiple GNN layers. In each layer, features are grouped and aggregated into matrices, resulting in an aggregated feature matrix \( Z_{\textit{agg}} \). The adjacency matrix for each layer is dynamically computed based on the feature similarity matrix \( S \):

\begin{equation}
S = Z_{\textit{agg}} \cdot Z_{\textit{agg}}^T
\label{eq:16}
\end{equation}

Self-loops are introduced by adding the identity matrix \( I \) to \( S \), and the modified adjacency matrix \( A \) is normalized as:

\begin{equation}
A = S + I, \quad \tilde{A} = \tilde{D}^{-0.5} \cdot A \cdot \tilde{D}^{-0.5}
\label{eq:17}
\end{equation}

Here, \(\tilde{D}\) is the diagonal degree matrix of \(A\), with each diagonal element \(\tilde{D}_{ii}\) being the sum of the \(i\)-th row of \(A\).

During forward propagation, each DGNN layer processes the inputs using a normalized adjacency matrix \( \tilde{A}_i \), a weight matrix \( W_i \), and a bias vector \( b_i \) for graph convolution. The outputs for each layer are calculated as:

\begin{equation}
\begin{cases}
    H_1 = \textit{ReLU}(\tilde{A}_1 \cdot Z_{\textit{agg}} \cdot W_1 + b_1) \\
    H_i = \textit{ReLU}(\tilde{A}_i \cdot H_{i-1} \cdot W_i + b_i), \quad i \in [2, n-1] \\
    H_n = \textit{ReLU}(\tilde{A}_n \cdot H_{n-1} \cdot W_n + b_n)
\end{cases}
\label{eq:18}
\end{equation}

Here, \( n \) denotes the total number of layers in the DGNN, which is set to \(3\) in our study.

The final \(\textit{Output}\) is computed through a sequence of operations including batch normalization \(\mathcal{F}_{\textit{bn}}(\cdot) \), dropout \(\mathcal{F}_{\textit{dropout}}(\cdot) \), and softmax activation \(\Phi_{\textit{softmax}} (\cdot)\), structured as:
\begin{equation}
\textit{Output} = \Phi_{\textit{softmax}}(\mathcal{F}_{\textit{dropout}}(\Gamma(\mathcal{F}_{\textit{bn}}(H^{(L)}))))
\label{eq:19}
\end{equation}
where \( \Gamma(\cdot) \) denotes the flattening operation, ensuring that the network output is appropriately structured for subsequent processing or analysis.

Finally, the procedure of AT-DGNN can be summarized in Algorithm~\ref{algorithm1}.

\begin{algorithm}[ht]
\caption{AT-DGNN}
\label{algorithm1}
\begin{algorithmic}[1]
\State \textbf{Input:} EEG data \( X_{i} \in \mathbb{R}^{E \times T} \), ground truth label \(y\)
\State \textbf{Output:} \( \textit{{pred}} \), the prediction of AT-DGNN
\State Initialization;

\Statex \Comment{Feature Extraction}
\For{\(i \leftarrow 1\) to \(3\)}
    \State get \(i\)th temporal kernel size by \eqref{eq:5};
    \State get \(Z^{i}_{\textit{temp}}\) by \eqref{eq:6} using \(X_{i}\) as input;
\EndFor
\State get \(Z_{T}\) by \eqref{eq:7};
\State get the window size \(n\) and segment the data by \eqref{eq:8}:

\For{\(w \leftarrow 1\) to \(n\)}
    \State get window \(X_w \in \mathbb{R}^{B \times E \times W}\);
    \State normalize \(X_w\) using LayerNorm;
    \State process \(X_w\) with MHA to get \(X_w'\) by \eqref{eq:9};
    \State normalize \(X_w'\) using LayerNorm;
    \State apply TCN on \( X_w' \) to get \( Z_{\textit{tcn}}^w \) by \eqref{eq:10};
    \State add \( Z_{\textit{tcn}}^w \) to \( Z_{\textit{stacked}} \);
\EndFor

\State apply Fusion Conv to \(Z_{\textit{stacked}}\) to get \(Z_{\textit{fused}}\) by \eqref{eq:11};

\Statex \Comment{Graph-based Learning}
\State perform graph filtering and aggregation on each node by \eqref{eq:12} - \eqref{eq:15} to get \(Z^i_{\textit{local}}\);
\For{\(i \leftarrow 1\) to \(3\)}
    \State get DGNN output by \eqref{eq:18};
\EndFor
\State get \(\textit{pred}\) by \eqref{eq:19};
\State \textbf{Return} \( \textit{pred} \)
\end{algorithmic}
\end{algorithm}

\section{Experiments} 

\subsection{MEEG Dataset}

In studies on emotion induction, visual stimuli such as images, videos, and text are commonly employed in experiments. However, participants' responses to these stimuli may be influenced by their cultural backgrounds \cite{juslin2008emotional}. Considering that the auditory cortex exhibits emotion-specific functional connections with a wide array of limbic, paralimbic, and neocortical structures, suggesting a more extensive role in emotion processing than previously understood \cite{koelsch2014brain}, we opted to induce emotions in participants using music in the MEEG dataset. Under the guidance of music professors, we selected 20 lesser-known Western classical music clips by composers such as Shostakovich and Tchaikovsky. Each clip lasts one minute and is characterized by distinct levels of valence and arousal. To minimize emotional carryover, a 15-second pause was introduced between clips \cite{2020Recognizing}. The study involved 32 students from Shandong University, aged between 20 and 25 years, who had no formal music education and were unfamiliar with Western classical music. Each participant signed an informed consent form prior to the study. This experimental design aimed to minimize the influence of social and cultural backgrounds on emotional responses, ensuring that changes in emotions were solely due to musical stimuli, thereby enhancing the accuracy of the experimental results. EEG data were collected using a 32-channel BCI device NeuSen.W32 from Neuracle Tech, employing the same electrode channels as in the DEAP dataset \cite{koelstra2011deap}. The data were sampled at a frequency of 1000 Hz. The EEG data were annotated based on the arousal and valence of the music during various stages of the experiment.

In contrast to the DEAP dataset, the MEEG dataset reduces experimental bias by rigorously selecting participants and musical stimuli, minimizing subjective factors and improving the effectiveness of emotional elicitation.

\subsection{Experiment Settings}

To rigorously evaluate the model's performance, a nested cross-validation approach is employed, featuring a trial-wise $10$-fold cross-validation in the outer loop and a $4$-fold cross-validation in the inner loop, as suggested by Varma \cite{varma2006bias}. This stratified sampling technique ensures robustness and generalization of the model by assessing its accuracy and reliability across diverse samples.

Additionally, a two-stage training strategy within the inner loop optimizes the utilization of training data. Initially, the best model identified from the $k$-fold cross-validation is saved as a preliminary candidate. This model is subsequently refined with the aggregated data from all $k$ folds, fine-tuned at a lower learning rate to avoid overfitting, and further trained for a maximum of 20 epochs or until it reaches 100\% training accuracy, ensuring precise calibration. Importantly, test data is excluded from the training to preserve evaluation integrity.

The integration of two-stage training with $10$-fold cross-validation minimizes variability in assessment results, providing a comprehensive and reliable evaluation framework suitable for diverse research and application domains.

\subsection{Implement Details} 

The model was implemented using PyTorch \cite{paszke2019pytorch} library. 

Cross-entropy loss was chosen as the objective function to guide the training process. The training was divided into two stages, with the first stage capped at 200 epochs and the second at 20 epochs. To reduce training time and prevent overfitting, early stopping was implemented. For the attention module, the window size was set to half the $f_{s}$. The kernel sizes for the time learner were set to 100, 50, and 25. Training was optimized using the Adam optimizer, starting with an initial learning rate of $1e-3$, which was reduced by a factor of 10 during the second stage. For more information on the data processing, model implementation details, and accessing the dataset, please refer to our GitHub repository \textit{https://github.com/xmh1011/AT-DGNN}.

\begin{table*}[ht]
\centering
\caption{Comparison of ACC and F1 scores for the MEEG dataset and DEAP dataset under trial-wise 10-fold cross-validation}
\label{table1}
\begin{threeparttable}
\begin{tabular}{c|l@{\hspace{10pt}}l|l@{\hspace{10pt}}l|l@{\hspace{10pt}}l|l@{\hspace{10pt}}l}
\toprule
\multirow{2}{*}{\textbf{Method}} & \multicolumn{2}{c|}{\textbf{MEEG Arousal}} & \multicolumn{2}{c|}{\textbf{MEEG Valence}} & \multicolumn{2}{c|}{\textbf{MEEG Emotion}} & \multicolumn{2}{c}{\textbf{DEAP Emotion}} \\
\cmidrule(lr){2-3} \cmidrule(lr){4-5} \cmidrule(lr){6-7} \cmidrule(lr){8-9}
& ACC (\%) & F1 (\%) & ACC (\%) & F1 (\%) & ACC (\%) & F1 (\%) & ACC (\%) & F1 (\%) \\
\midrule
EEGNet         & 75.47$^{**}$ & 75.54$^{**}$ & 73.66$^{**}$ & 73.33$^{**}$ & 74.57 & 74.44 & 60.36 & 64.64 \\
DeepConvNet    & 79.35$^{**}$ & 78.43$^{**}$ & 72.38$^{**}$ & 73.03$^{**}$ & 75.87 & 75.73 & 60.41 & 63.80 \\
ShallowConvNet & 80.96$^{**}$ & 80.87$^{**}$ & 79.77$^{**}$ & 80.22$^{**}$ & 80.37 & 80.55 & \textbf{61.01} & 64.44 \\
TSception      & 79.23$^{**}$ & 81.30$^{*}$  & 78.34$^{**}$ & 81.18$^{**}$ & 78.79 & 81.24 & 60.42 & 64.15 \\
EEG-TCNet      & 76.36$^{**}$ & 73.79$^{**}$ & 69.72$^{**}$ & 54.28$^{***}$ & 73.04 & 64.04 & 56.11 & 52.50 \\
TCNet-Fusion   & 76.72$^{**}$ & 76.95$^{**}$ & 75.88$^{**}$ & 75.42$^{**}$ & 76.30 & 76.20 & 59.33 & 63.94 \\
ATCNet         & 82.01$^{**}$ & 79.17$^{**}$ & 83.19$^{**}$ & 82.08$^{**}$ & 82.60 & 80.63 & 60.83 & 62.32 \\
DGCNN          & 82.12$^{**}$ & 81.83$^{**}$ & 82.72$^{**}$ & 82.40$^{**}$ & 82.42 & 82.12 & 60.36 & 61.27 \\
LGGNet-Fro     & 81.85$^{**}$ & 81.58$^{**}$ & 84.28$^{**}$ & 83.87$^{**}$ & 83.07 & 82.73 & 60.59 & 64.48 \\
LGGNet-Gen     & 82.15$^{**}$ & 82.15$^{**}$ & 84.53$^{**}$ & 84.38$^{**}$ & 83.34 & 83.27 & 60.96 & \textbf{65.41} \\
LGGNet-Hem     & 81.92$^{**}$ & 81.58$^{**}$ & 84.93$^{**}$ & 84.46$^{**}$ & 83.42 & 83.02 & 59.88 & 62.50 \\
\midrule
AT-DGNN-Fro    & 83.51$^{**}$ & 83.06$^{**}$ & 85.56$^{**}$ & 85.61$^{**}$ & 84.54 & 84.34 & 59.95 & 62.61 \\
\textbf{AT-DGNN-Gen} & \textbf{83.74}$^{**}$ & 84.56$^{**}$ & \textbf{86.01}$^{*}$ & \textbf{85.68}$^{*}$ & \textbf{84.88} & \textbf{85.12} & \textbf{60.55} & \textbf{63.83} \\
AT-DGNN-Hem    & 83.73$^{**}$ & \textbf{84.78}$^{**}$ & 84.89$^{**}$ & 85.35$^{**}$ & 84.31 & 85.07 & 59.36 & 62.16 \\
\bottomrule  
\end{tabular}
\begin{tablenotes}
\small
\item * indicating $std < 10\%$, ** indicating $std < 20\%$, *** indicating $std < 30\%$.
\item Emotion averages the ACC and F1 scores of both the arousal and valence dimensions.
\item The results for are derived from the optimal values among ten trials with distinct random seeds.
\end{tablenotes}
\end{threeparttable}
\end{table*}

\section{RESULTS AND DISCUSSION}

In this section, we compare the average ACC and F1 score of AT-DGNN on the MEEG dataset with CNN, TCN, and GNN-based SOTA methods in the BCI domain. The CNN-based methods include: EEGNet \cite{lawhern2018eegnet}, TSception \cite{ding2020tsception}, DeepConvNet and ShallowConvNet \cite{schirrmeister2017deep}. The TCN-based methods include: EEG-TCNet \cite{ingolfsson2020eeg}, TCNet-Fusion \cite{musallam2021electroencephalography} and ATCNet \cite{altaheri2022physics}. The GNN-based methods include: DGCNN \cite{song2018eeg} and LGGNet \cite{ding2023lggnet}. Due to the smaller size of the CNN-based models, the learning rate and the number of training epochs are reduced to avoid overfitting. For the other models, the parameters recommended by the authors are used. Additionally, ablation studies are conducted to reveal the contribution of each component within the AT-DGNN architecture. 

\subsection{Emotion Recognition}

As shown in Table~\ref{table1}, the AT-DGNN series, particularly the AT-DGNN-Gen model, significantly outperforms the previous state-of-the-art LGGNet series in both arousal and valence dimensions, with notable increases in F1 scores and accuracy. Specifically, the AT-DGNN-Gen model achieves the highest accuracy of 83.74\% in the arousal dimension, surpassing the LGGNet-Gen model by 1.59\%, and records the highest accuracy of 86.01\% in the valence dimension, marking a substantial improvement of 1.08\% over the LGGNet series. This improvement underscores the efficacy of our approach in capturing complex patterns within the MEEG dataset, which is known for its challenging and class-imbalanced nature.

Moreover, the AT-DGNN-Gen model exemplifies the superior performance of the AT-DGNN series, achieving robust results with an average accuracy of 84.88\% and an F1 score of 85.12\% across emotion dimensions. This highlights the effectiveness of our graph-based approach in handling complex emotion recognition tasks. Furthermore, consistent with the findings of LGGNet~\cite{ding2023lggnet}, the graph configuration \(G_g\) demonstrated superior performance in both AT-DGNN and LGGNet models compared to other graph definitions, indicating its optimal structure for capturing relevant features in emotion analysis.

The results also show a significant improvement over traditional CNN and TCN-based methods, with average gains in F1 scores and accuracy exceeding previous models by up to 7.16\% and 5.78\%, respectively. This highlights the enhanced capability of the DGNN approach in analyzing graph-structured data, offering more accurate and reliable emotion recognition. Additionally, the AT-DGNN exhibits a lower standard deviation compared to other models, indicating more balanced performance across various samples and superior generalization ability.

On both the MEEG and DEAP datasets, the AT-DGNN demonstrates strong generalization ability. All models achieve high accuracy on the MEEG dataset, with CNN and TCN models reaching 75\% or higher, and GNN models exceeding 80\%. Specifically, all models, including AT-DGNN, achieve significantly higher accuracy on the MEEG dataset than on the DEAP dataset, suggesting that music is more effective than video in evoking emotions and indicating the higher quality of the MEEG dataset. On the DEAP dataset, the AT-DGNN achieves an average accuracy of 60.56\% across four dimensions, close to other SOTA models and surpassing some of them, which demonstrates the AT-DGNN model's strong generalization capability across different datasets.

\subsection{Ablation Study}

\begin{table}[ht]
\centering
\caption{Ablation study on Emotion of MEEG using AT-DGNN-Gen.}
\label{tab:ablation_study}
\begin{threeparttable}
\begin{tabular}{cccc|c|c|c|c}
\toprule    
S & A & G & T & ACC(\%) & Changes(\%) & F1(\%) & Changes(\%) \\  
\midrule
\cmark & \xmark & \textbf{\textit{\fontsize{10}{13}\selectfont 3}} & \textbf{\textit{\fontsize{10}{13}\selectfont 3}} & 78.31 & -6.57 & 79.46 & -5.66 \\
\xmark & \cmark & \textbf{\textit{\fontsize{10}{13}\selectfont 3}} & \textbf{\textit{\fontsize{10}{13}\selectfont 3}} & 80.89 & -3.99 & 81.16 & -3.96 \\
\midrule 
%一层图神经网络
\cmark & \cmark & \textbf{\textit{\fontsize{10}{13}\selectfont 1}} & \textbf{\textit{\fontsize{10}{13}\selectfont 3}} & 81.91 & -2.97 & 81.81 & -3.31 \\
%两层动态图神经网络
\cmark & \cmark & \textbf{\textit{\fontsize{10}{13}\selectfont 2}} & \textbf{\textit{\fontsize{10}{13}\selectfont 3}} & 82.94 & -1.94 & 82.52 & -2.60 \\ 
%四层动态图神经网络
\cmark & \cmark & \textbf{\textit{\fontsize{10}{13}\selectfont 4}} & \textbf{\textit{\fontsize{10}{13}\selectfont 3}} & 83.17 & -1.71 & 83.48 & -1.64 \\
% 无图神经网络
\cmark & \cmark & \xmark & \textbf{\textit{\fontsize{10}{13}\selectfont 3}} & 83.05 & -1.83 & 83.89 & -1.23 \\
\midrule 
% 一层时间学习器
\cmark & \cmark & \textbf{\textit{\fontsize{10}{13}\selectfont 3}} & \textbf{\textit{\fontsize{10}{13}\selectfont 1}} & 82.95 & -1.93 & 82.84 & -2.28 \\
%两层时间学习器
\cmark & \cmark & \textbf{\textit{\fontsize{10}{13}\selectfont 3}} & \textbf{\textit{\fontsize{10}{13}\selectfont 2}} & 82.60 & -2.28 & 81.91 & -3.21 \\ 
%四层时间学习器
\cmark & \cmark & \textbf{\textit{\fontsize{10}{13}\selectfont 3}} & \textbf{\textit{\fontsize{10}{13}\selectfont 4}} & 82.89 & -1.99 & 83.43 & -1.69 \\ 
%不实现时间学习器
\cmark & \cmark & \textbf{\textit{\fontsize{10}{13}\selectfont 3}} & \xmark & 74.64 & -10.24 & 74.84 & -10.28 \\
%三层时间学习器，三层动态图神经网络
\midrule 
\cmark & \cmark & \textbf{\textit{\fontsize{10}{13}\selectfont 3}} & \textbf{\textit{\fontsize{10}{13}\selectfont 3}} & \textbf{84.88} & - &
\textbf{85.12} & - \\
\bottomrule  
\end{tabular}
\begin{tablenotes}
\small
\item S: Sliding window segmentation.
\item A: Multi-head attention mechanism.
\item G: Graph neural network.
\item T: Temporal learner.
\item \cmark: Keep the component.
\item \xmark: Exclude the component.
\item \textit{1}, \textit{2}, \textit{3}, \textit{4}: The number of DGNN layers or temporal learner.
\item Changes: Compared with the baseline AT-DGNN-Gen.
\end{tablenotes}
\end{threeparttable}
\end{table}

As shown in Table~\ref{tab:ablation_study}, the ablation study systematically evaluates the contributions of key components in the AT-DGNN-Gen model, including sliding window segmentation (S), multi-head attention mechanism (A), graph neural network layers (G), and temporal learner layers (T). The baseline model, incorporating all components with three layers each for G and T, achieved the highest ACC of 84.88\% and F1 score of 85.12\%.

Excluding the multi-head attention mechanism (A) caused a significant performance drop, with ACC decreasing by 6.57\% and F1 score by 5.66\%. Similarly, removing the sliding window segmentation (S) resulted in reductions of 3.99\% in ACC and 3.96\% in F1 score. These results confirm the essential roles of S and A in feature extraction and temporal modeling.

Varying the number of G layers showed that using three layers yielded optimal results, while deviations caused performance declines. Increasing G layers to four resulted in a minor ACC drop of 1.71\%, while reducing to one layer led to a larger decline of 2.97\%. Similar trends were observed for T layers, where three layers provided the best balance, and exclusion of T caused a substantial drop of 10.24\% in ACC and 10.28\% in F1 score, highlighting its critical role in capturing temporal dynamics.

Overall, the ablation study demonstrates that each component of the AT-DGNN-Gen model contributes significantly to its performance. The sliding window segmentation and multi-head attention mechanism are crucial for effective feature extraction, while the optimal number of layers for the graph neural network and temporal learner is three, balancing model complexity and performance. These findings substantiate the design choices of our model and underscore the importance of the synergistic operation of its components in enhancing emotion recognition from EEG data. The results emphasize how the model's architecture is essential for capturing the complex spatiotemporal patterns inherent in EEG signals, thereby advancing computational neuroscience analysis.

\section{Conclusion}

This study introduces the MEEG dataset and the AT-DGNN framework, significantly advancing EEG-based emotion recognition. The MEEG dataset utilizes music to induce emotional states, providing a unique and critical resource for analyzing brain responses to emotional stimuli. The AT-DGNN framework captures the complex temporal dynamics of brain activity, enhancing emotion recognition accuracy beyond existing SOTA methods. Moreover, the MEEG dataset displays emotional states of subjects with greater precision, enabling SOTA methods to achieve higher accuracy and establishing a more robust benchmark for evaluating EEG-based emotion analysis models. These advancements propel BCI technology forward and facilitate new research into the relationships between music, emotion, and brain activity.

\section{ACKNOWLEDGEMENTS}
This paper is supported by the Shenzhen Fundamental Research Program under Grant JCYJ20230807094104009.
and Key Lab of Information Network Security, Ministry of Public Security.
\small
\bibliographystyle{IEEEtran}
\bibliography{cite.bib}

\end{document}